\documentclass[preprint,showpacs,preprintnumbers,amsmath,amssymb]{revtex4}

\usepackage{graphicx}
\usepackage{dcolumn}
\usepackage{bm}

\begin{document}

\vspace{5mm}

\newcommand{\goo}{\,\raisebox{-.5ex}{$\stackrel{>}{\scriptstyle\sim}$}\,}
\newcommand{\loo}{\,\raisebox{-.5ex}{$\stackrel{<}{\scriptstyle\sim}$}\,}

\title{Formation of hypernuclei in heavy-ion collisions around the threshold 
energies.}

\author{A.S.~Botvina$^{1,2}$, K.K.~Gudima$^{1,3}$,
J.~Steinheimer$^{1}$, M.~Bleicher$^{1}$, J.~Pochodzalla$^{4}$}

\affiliation{$^1$Frankfurt Institute for Advanced Studies and ITP J.W. Goethe 
University, D-60438 Frankfurt am Main, Germany} 
\affiliation{$^2$Institute for Nuclear 
Research, Russian Academy of Sciences, 117312 Moscow, Russia} 
\affiliation{$^3$Institute of Applied Physics, Academy of Sciences of Moldova, 
MD-2028 Kishinev, Moldova} 
\affiliation{$^4$ Helmholtz-Institut Mainz and Institut f{\"u}r Kernphysik, 
J.Gutenberg-Universit{\"a}t Mainz, D-55099 Germany}

\date{\today}

\begin{abstract}

In relativistic ion collisions there are excellent opportunities to produce 
and investigate hyper-nuclei. We have systematically studied the formation of 
hypernuclear spectator residues in peripheral heavy-ion collisions with the 
transport DCM and UrQMD models. The hyperon capture was calculated within the 
potential and coalescence approaches. We demonstrate that even at the 
beam energies 
around and lower than the threshold for producing $\Lambda$ hyperons in binary 
nucleon-nucleon interactions a considerable amount of hypernuclei, 
including multi-strange ones, can be produced. This is important for 
preparation of new experiments on hypernuclei in the wide energy range. 
The uncertainties of the predictions are investigated within the models, 
and the comparison with the strangeness production measured in experiments 
is also performed. 

\end{abstract}


\maketitle

\section{Introduction}

In nuclear reactions the detailed studies of the three-flavor 
processes, including $u$, $d$ and $s$ quarks, will be mandatory to 
develop fundamental nuclear theories of hadrons and nuclei, as well as 
the large spectacular cosmic objects like neutron stars. 
Having a core with supra-nuclear densities and a crust with 
sub-nuclear densities, these stellar objects merge all aspects of nuclear 
physics. Baryons with strangeness embedded in 
the nuclear environment, i.e., hypernuclei, are the only available tool to 
approach the many-body aspect of the strong interaction at low energies. 
Hypernuclei are formed when hyperons ($Y=\Lambda,\Sigma,\Xi,\Omega$) 
produced in high-energy interactions are captured by nuclei. They live 
significantly longer than the typical reaction times, therefore, they 
can serve as a  tool to study the hyperon--nucleon and hyperon--hyperon 
interactions. 
The investigation of hypernuclei is a very progressing 
field of nuclear physics, since it provides complementary methods to 
improve traditional nuclear methods and open new horizons for studying 
particle physics and nuclear astrophysics 
(see, e.g., \cite{Ban90,Sch93,Gre96,Has06,Sch08,Gal12,Buy13,Hel14} 
and references therein).

Recently very encouraging results on hypernuclei 
come from experiments with relativistic ion collisions. 
Many experimental collaborations (e.g., STAR at RHIC \cite{star}; 
ALICE at LHC \cite{alice}; PANDA \cite{panda}, FOPI, HADES, CBM 
\cite{Hades,Vas15}, and 
HypHI, Super-FRS, R3B at FAIR \cite{saito-new,super-frs,Rap13}; 
BM@N and MPD at 
NICA \cite{nica}) have started or plan to investigate hypernuclei and 
their properties in reactions induced by relativistic hadrons and ions. 
The limits in isospin space, unstable nuclear states, multiple strange 
nuclei and precision lifetime measurements are unique topics of these 
fragmentation reactions. 

It is important in this respect to note that the very first experimental 
observation of a hypernucleus was obtained in the 1950-s in reactions of 
nuclear multifragmentation induced by cosmic rays \cite{Dan53}. Recently 
a remarkable progress was made in investigation of the multifragmentation 
reactions associated with relativistic heavy-ion collisions (see, e.g., 
\cite{Bon95,Xi97,Sch01,Ogu11} and references therein). This gives us an 
opportunity to apply well known theoretical methods adopted for description 
of these reactions also for production of hypernuclei \cite{Bot07,Das09}.

Specially, we emphasize a possibility to form hypernuclei 
in the fragmentation processes in peripheral collisions. 
The insight into mechanisms of such processes will provide access to the EoS 
of hyper-nuclear matter and explain the phase 
transition phenomena at low temperature. 
As already discussed \cite{Buy13,Bot11} in these reactions one can get a very 
broad distribution of produced hypernuclei including the exotic ones 
and with extreme isospin. This can help to investigate the structure 
of nuclei by extending the nuclear chart into the strangeness sector 
\cite{Ban90,Sch93,Gre96,Has06}. In addition, 
complex multi-hypernuclear systems incorporating more than two hyperons 
can be created in such energetic nucleus-nucleus collisions, and this 
may be the only conceivable method to go even beyond $|s|$=2.

It was demonstrated in the previous works \cite{Bot13,Bot15} that the yields 
of single hypernuclei originating from the spectator residues in 
peripheral ion collisions will saturate with energies above 3--5 A GeV 
(in the laboratory frame). Therefore, the accelerators of moderate 
relativistic energies can be used for the intensive studies of 
hypernuclei. The subthreshold production of hyperons 
becomes possible in these reactions down to the energies of $\sim$1 A GeV. 
At the laboratory energies of ions around 1--2 A GeV 
the detection of hypernuclei can become very effective and this give an 
advantage despite of their smaller yields in comparison with high energy 
beams. For example, the novel experimental set-ups, like FRS/Super-FRS 
\cite{Rap13,aumann,frs}, 
can be effectively used for separation of nuclei with energies less than 
2 A GeV, that essentially extends opportunities for their investigation. 
This gives chances to measure many new exotic hypernuclei. 
The acceptance of other modern detectors, like CBM at GSI/FAIR \cite{Vas15} 
or STAR at RHIC \cite{star} allows to register particles coming mostly 
from the kinematic region around midrapidity. Therefore, the decreasing 
of the beam energy will increase considerably the probability for fragments 
produced in the target/projectile region to enter the detection domain. 
Since the formation of large hypernuclei is shifted toward the 
target/projectile rapidities \cite{Bot15}, this can open possibilities 
to form novel hypernuclear states. Another research direction is related to 
producing multi-strange hyper-fragments which may require a higher beam 
energy. This needs a systematic theoretical investigation of double- and 
multi-strange fragment yields at least at beam energies up to $\sim$10 
A GeV in order to understand if any saturation phenomenon can be observed. 
The light multi-strange clusters can be measured with the high precision 
detectors, for example, by CBM collaboration \cite{Vas15}. The aim of 
this theoretical work is to investigate the hypernuclei production in 
detail the region of the beam energies of 1--10 A GeV, 
including the comparison with available experiments and the parameter 
dependence of the results, 
in order to 
provide the future experiments with reliable predictions of these reactions. 


\section{Models for production of hypernuclei at relativistic collisions}

We recall shortly the mechanisms for producing hypernuclei which 
were discussed previously: 
The formation processes of hypernuclei are apparently different in central 
and peripheral ion collisions. There are indications that in central 
collisions of very high energy the coalescence mechanism, which 
assembles light hyper-fragments from the produced hyperons and
nucleons (including anti-baryons), is essential 
\cite{star,alice,ygma-nufra,camerini-nufra}.
Thermal models suggest also that only the lightest clusters, with 
mass numbers A$\loo$4, can be noticeably produced in this way because 
of the very high temperature of the fireball (T$\approx$160 MeV) 
\cite{And11,Ste12}.
On the other hand, it was claimed long ago that the absorption 
of hyperons in the spectator regions
after peripheral nuclear collisions is a promising way to produce 
hypernuclei \cite{Bot11,Wak88,Cas95,giessen}. 
The special reactions associated with these processes, e.g., the 
hyper-fission, were under investigation too \cite{Arm93,Ohm97}. 
An important feature of peripheral collisions is that large pieces of 
nuclear matter around normal nuclear density at low temperature 
can be created in contrast to the highly-excited nuclear matter 
at mid-rapidity. Nucleons from the overlapping parts of the projectile 
and target (participant zone) interact strongly among themselves 
and with other hadrons produced in primary and secondary collisions. 
Nucleons from the non-overlapping parts do not interact intensively, and 
they form residual nuclear systems, which we call spectator residues. 
We remind that these residues are formed during first 30--60 fm/c 
after starting the collision, when energetic hadron-nucleon interactions 
inside nuclei cease and the remaining nucleons do not escape the 
nucleus potential \cite{Bon95,Bot11}. 
The nuclear system evolves toward thermalization in this case. 
It is well established that low excited spectator 
residues (T$\loo$5-6 MeV) are produced in such reactions 
\cite{Bon95,Xi97,Sch01,Poc97}. 
The production of hyperons is associated with nucleon-nucleon collisions, 
e.g.,  p+n$\rightarrow$n+$\Lambda$+K$^{+}$, or collisions of secondary 
mesons with nucleons, e.g., $\pi^{+}$+n$\rightarrow \Lambda$+K$^{+}$. 
Strange particles may be produced in the participant zone, however, the 
particles can re-scatter and undergo secondary interactions. As a result 
the produced hyperons populate the whole momentum space around the 
colliding nuclei, including the vicinity of nuclear spectators, 
and can be captured by the spectator residues. General 
regularities of the decay of such hyper-residues into hyper-fragments 
can be investigated with statistical models (e.g., generalized 
Statistical Multifragmentation Model SMM \cite{Bot07,Das09}), which were 
previously applied for description of normal fragments in similar processes 
with great success \cite{Bon95,Xi97,Sch01,Ogu11}. 

The theoretical predictions of strangeness and hyperon production 
in hadron and ion reactions can be performed with various
dynamical models employing similar general assumptions on the
hadron transport in nuclei but with different methods of solution of the
kinetic equations. In addition, the models can also be different 
(especially at high energy) in the description of elementary hadron-hadron 
interactions and production of new particles. Previously we have investigated 
the model-dependence of the results \cite {Bot11,Bot15}. 
At relatively low-energy elementary hadron collisions (less than 1--3 GeV 
in the laboratory frame) the models use usually some 
approximations for the reaction channels supported by the analysis 
of available experimental data. However, at higher energies, where 
hyperon formation probability is large, theoretical evaluations are mostly 
used. For example, the Dubna Cascade Model (DCM) \cite{Bot11,Bot13,Ton90} 
involves the quark gluon string model (QGSM). The Ultrarelativistic Quantum
Molecular Dynamics (UrQMD) model \cite{Bas98,Ble99} has adopted the string
formation and its fragmentation according to the PYTHIA model for hard
collisions. 
In particular, the current versions of DCM and UrQMD include up to 
70 baryonic species (including their anti-particles), as well as up to 
40 different mesonic species, which are involved in binary interactions. 
The Lund FRITIOF string model (including PYTHIA) is used in
the Hadron String Dynamics model (HSD) \cite{Cas08}, however, for 
simulations including in-medium self-energies of particles. 
We have shown that at high energy the difference between these models 
have a moderate 
influence (within the factor of two) on the yield of hypernuclei, that can 
be considered as an uncertainty of their prediction \cite{Bot15}. 
The capture of produced $\Lambda$ hyperons by nuclear spectator residues 
can be easily obtained within the potential criterion \cite{Bot11}: 
It takes place if a hyperon 
kinetic energy in the rest frame of the residue is lower than the 
attractive potential energy generated by neighbouring nucleons, i.e., 
the hyperon potential, which is around 30 MeV in matter at normal nuclear 
density $\rho_0 \approx 0.15 fm^{-3}$. 
The variation (mostly, decreasing) of the nuclear density is taken into 
account during the hadron cascade development in nuclei and the hyperon 
capture potential varies correspondently \cite{Bot11}. 
The coalescence criterion \cite{Ste12}, which uses 
the proximity of baryons in momentum and coordinate space, 
is consistent with the potential one. 
A generalization of the coalescence model \cite{Bot15}, the 
coalescence of baryons (CB), can be applied after the dynamical stage 
described, for example, by DCM, UrQMD, and HSD models. 
In such a way it is possible to form primary 
fragments of all sizes, from the lightest nuclei to the heavy residues, 
including hypernuclei within the same mechanism. 
We have found previously \cite{Bot15} that the optimal time for applying the 
coalescence (as the final state interaction) is around 40--50 fm/c after 
starting the heavy-ion collisions, when the rate of individual inelastic 
hadron interactions decreases very rapidly. A variation of the time within 
this interval leads to an uncertainty in the yield around 10$\%$ for a fixed 
coalescence parameter. This is essentially smaller than the uncertainty in 
the coalescence parameter itself. 
It is important that the calculations are performed on the event-by-event 
basis, like the experimental data are obtained. 
The following break-up of excited primary fragments can be described 
with the statistical models \cite{lorente,Bot07,Bot13} by using the same 
Monte-Carlo method, which allows to keep information on each produced 
particle. The advantage of this hybrid procedure is the possibility 
to predict the correlations of yields of hypernuclei, including their sizes, 
with the rapidity, and with other produced particles. 

In this paper we concentrate on the transport approaches and the capture 
of $\Lambda$ hyperons. In particular, we show new systematic calculations 
with DCM and UrQMD models for various target/projectiles at relevant 
energies, as well as the comparison with available experimental data 
on strangeness production. 
We demonstrate also the sensitivity of the hyper-fragment yields 
to the parametrization of the hyperon production and its capture. 
We believe that in this way one can realistically estimate 
the primary hyper-fragments yields that is important 
for the planning of future experiments.


\begin{figure}[tbh]
\includegraphics[width=0.6\textwidth]{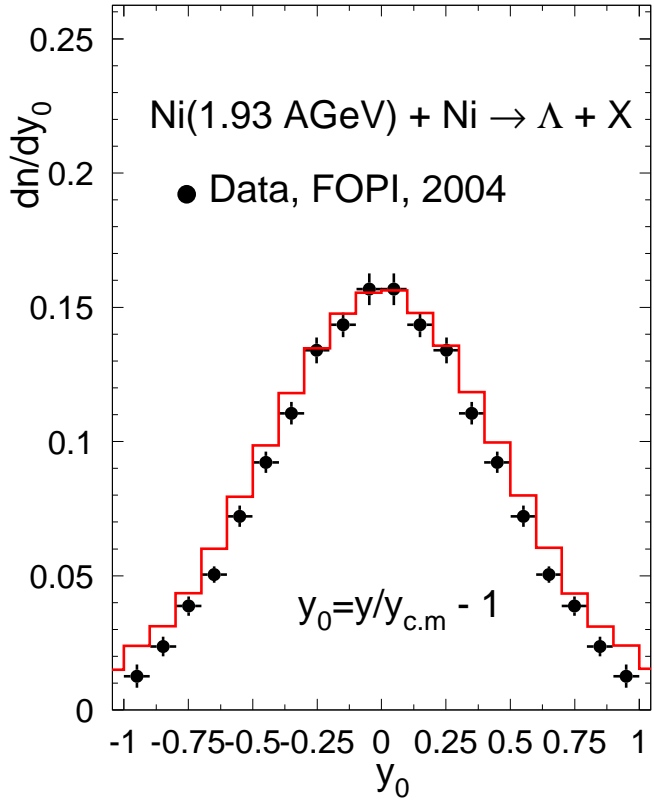}
\caption{\small{ (Color online)
Rapidity distribution of $\Lambda$ hyperons as  measured by FOPI 
collaboration \cite{FOPI04} in comparison with DCM calculations. 
The Ni + Ni reaction at 1.93 A GeV is analyzed for central 
and semi-central events. 
}}
\label{fig1}
\end{figure}

\begin{figure}[tbh]
\includegraphics[width=0.6\textwidth]{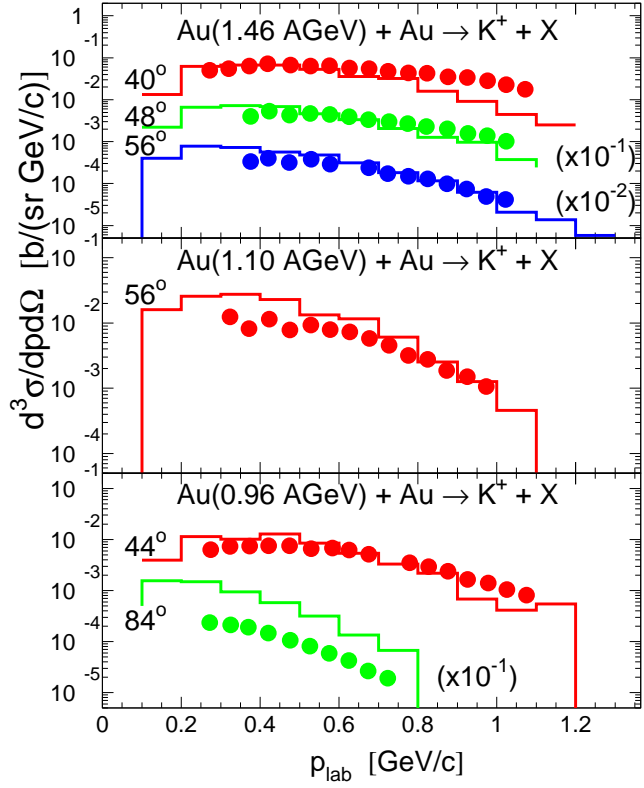}
\caption{\small{ (Color online)
Double differential cross sections as functions of the laboratory particle 
momenta for the production of K$^{+}$ mesons in the gold-on-gold collisions 
at subthreshold energies under different angles. 
The experimental data (solid circles) are taken from Ref.~\cite{PRL01}. 
The DCM calculations (solid histograms) are integrated over all 
impact parameters to meet the experimental conditions. 
The energies and angles in the laboratory system are given in the panels. 
The scaling factors 
are given in the brackets. 
}}
\label{fig2}
\end{figure}

\begin{figure}[tbh]
\includegraphics[width=0.6\textwidth]{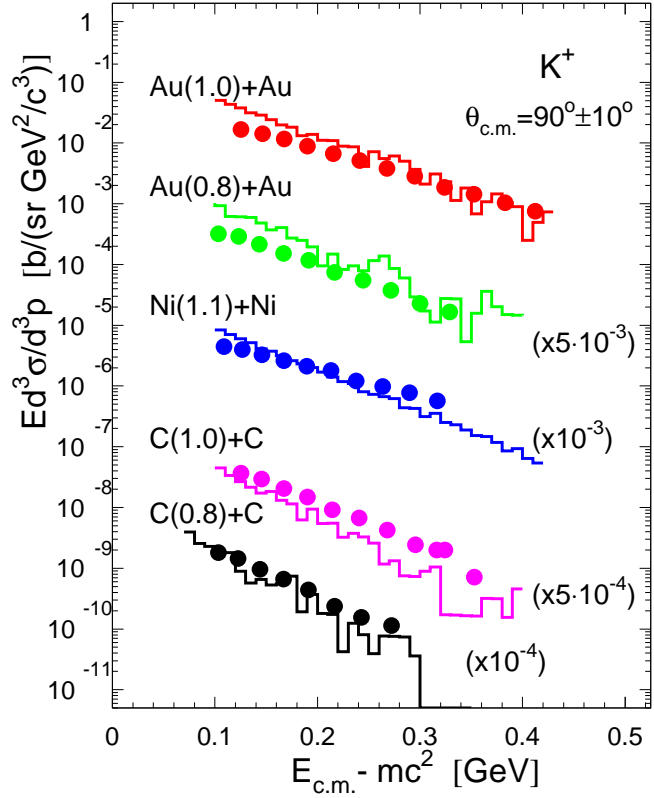}
\caption{\small{ (Color online)
Invariant cross sections for the production of K$^{+}$ mesons in the 
center-of-mass system versus their kinetic energy under the angle of 90 
degree, in the gold, nickel and carbon symmetric ion collisions. 
The experimental data (solid symbols) are taken from Ref.~\cite{PRC07}. 
The DCM calculations are given by histograms. 
The energies in the laboratory system (in A GeV) and the 
scaling factors are shown in the panels. 
}}
\label{fig3}
\end{figure}

\begin{figure}[tbh]
\includegraphics[width=0.6\textwidth]{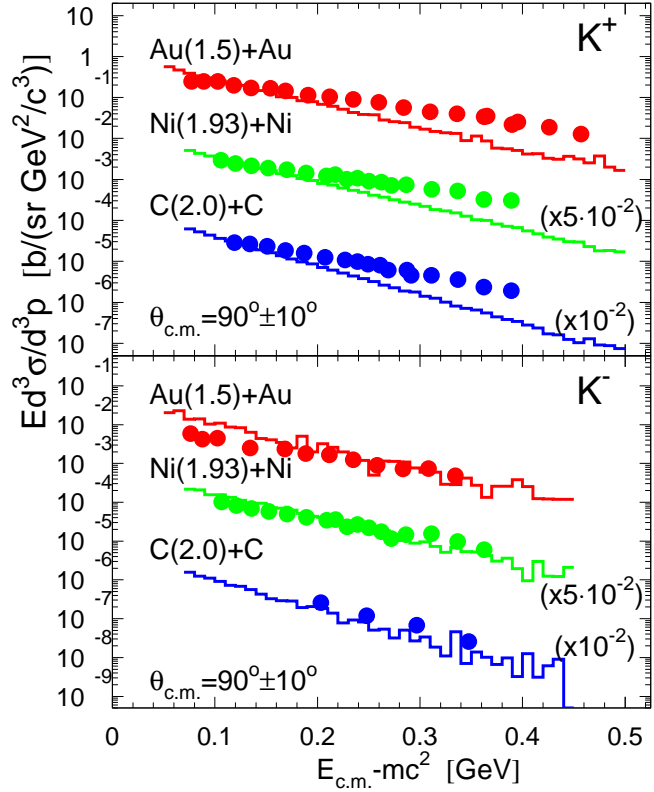}
\caption{\small{ (Color online)
The same as in Fig.~3, however, in the collisions at higher energies, 
and for K$^{+}$ and K$^{-}$ mesons (top and bottom panels). 
}}
\label{fig4}
\end{figure}

\begin{figure}[tbh]
\includegraphics[width=0.6\textwidth]{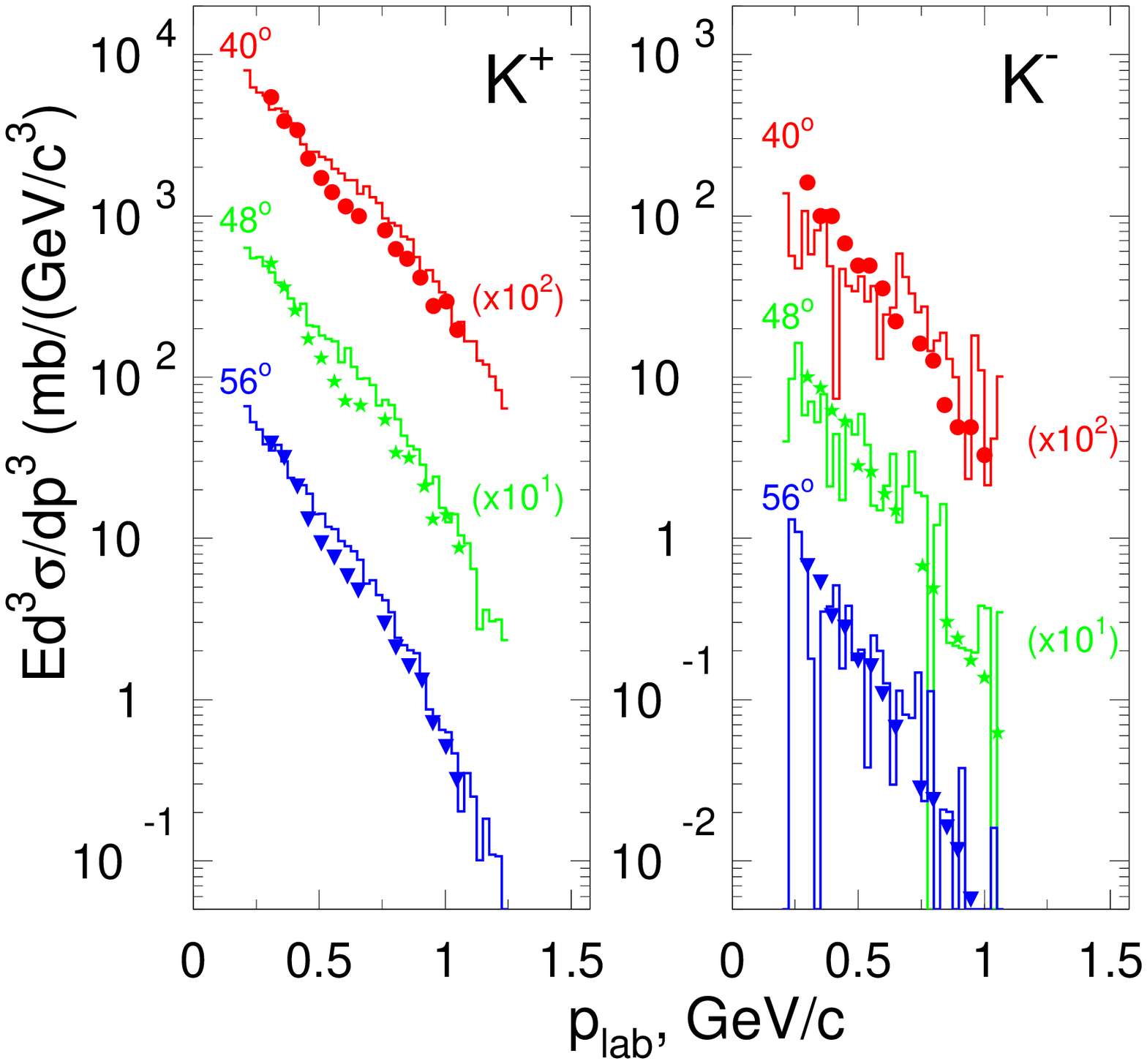}
\caption{\small{ (Color online)
Invariant cross sections for the production of K$^{+}$ and K$^{-}$ mesons 
in reactions induced by protons with the energies of 3.5 GeV on 
the gold target versus their momenta under several angles. All values are 
in the laboratory system, see notations on the panels. The symbols are 
experimental data taken from Ref.~\cite{PRL06}, 
the DCM calculations are histograms. 
}}
\label{fig5}
\end{figure}

\section{Formation of hyperons and strange particles} 

The transport models were used successfully for description of strangeness 
production (see, e.g., Refs.~\cite{Bot11,Bra04,Har12,Fuc01}). However, 
there are only few experimental data concerning the hyperon production. 
Some of them were analyzed in the previous works \cite{Bot11}. In Fig.~1 
we show the comparison of DCM with the $\Lambda$-hyperon rapidity 
distributions measured by FOPI (GSI) collaboration for central and 
semi-central events (the estimated impact parameters are less than 5.5 fm). 
One can see a rather good agreement, which was obtained when the processes 
involving secondary interactions of produced particles were included 
\cite{Cas95,Bot11}. 
In the rapidity region of the projectile/spectators, one can note 
a slight surplus of the hyperons in the calculations. This should be tested 
with future measurements when the experimental efficiency will be improved 
at these rapidities. 

The above projectile energy is still larger than the threshold for $\Lambda$ 
production in the nucleon-nucleon collision ($E_{thr} \approx 1.6$ A GeV). 
On the other hand the subthreshold production is possible, because of the 
Fermi motion of nucleons in colliding nuclei and secondary rescattering 
processes. In order to 
verify the calculations at energies lower than $E_{thr}$ we could look at 
the reaction products which accompany the hyperon production in this 
case. In particular, the channels with the K$^{+}$ formation,  are 
dominating here. Therefore, in Fig.~2 and Fig.~3 we analyze the yields 
of positive kaons at the subthreshold energies. 
One see a quite reasonable agreement 
of the DCM transport calculations with the KAOS experimental data on the 
differential spectra at various angles. We should take into account that 
namely these spectra are used for the final fit and evaluation 
of the total kaon yields in the experiments. One can see a slight 
overestimation of the kaon production in the model under large angles 
(in the backward direction). However, they are responsible for a small part 
of the yield. Unfortunately, it was not possible to detect kaons with 
very low momenta and this leads to an uncertainty in the estimate of the 
total yield. However, we believe that within the model 
we can give a reasonable evaluation of the strangeness production 
in the subthreshold region. It is sufficient for the preparation of 
experiments on strange particles and fragments, if one can detect 
such particles (via products of their decay) 
with the cross sections of around nanobarns. 

If we analyze the laboratory energies around and above the threshold, then 
the model predictions reproduce the experimental data better. 
This can be seen from Figs.~4 and 5. An underestimation of the positive 
kaons of high energy with DCM (Fig.~4) should have a minor effect, and it 
is just indicating that the number of $\Lambda$ hyperons, which accompany 
the kaons, may be even larger. Though at this energy range the kaons were 
under selective study with other transport models (for example, see 
\cite{Bra04,Har12} and references therein) the mechanism of their 
formation is still under discussion. We believe that the available 
direct hyperon distributions, shown here in Fig.~1 and in Ref.~\cite{Bot11}, 
should be analyzed within other models too. 
As seen from our analysis all kaon spectra can be reasonably 
described within DCM transport approach and this is a good justification 
for extending the model predictions to other strange particles. In our 
opinion it would be useful to perform a cross-comparison of different 
models on strangeness production around the threshold energy for clarifying 
the physics behind. 
Having in mind that controlling 
verifications of the transport codes are still necessary, we think 
we can afford a reasonable estimate of the hyper-fragment yields 
with the determination of the yield uncertainty also. In this case the 
effect of the hyperon capture by nucleons and clusters can be evaluated 
with help of additional model parameters.

\section{Production of hyper-residues} 

It was discussed that the hyperon capture can be described by both the 
potential and the coalescence approach \cite{Bot11,Bot15}. The coalescence 
is very popular and it exists on the market in different modifications, 
see, for example, its application for the hyperon capture within other 
transport approaches \cite{giessen,Gai14,LeFev16}. The connection of the 
potential and coalescence capture conceptions was demonstrated previously. 
In particular, the momentum distribution of the hyperon 
captured in the nuclear potential well reminds a step-like function, 
see Fig.~10 in Ref.~\cite{Bot11}. This can be approximated with 
the coalescent capture of hyperons if their momenta (or velocities) relative 
to other nucleons less than a certain value. It is important 
to compare the capture probability within the two approaches. 
Our understanding is that this capture is a fast process, which 
happens during the time around few tens fm/c from the beginning of the 
reaction. As a result, the most produced hyper-clusters should be excited 
from low excitations up to few MeV per nucleons. They will decay afterwards 
during a very prolonged time ($\sim 10^{2}-10^{4}$ fm/c). It is well known 
that the secondary decay involves many-particle correlations which are usually 
not included in transport models. Such decay processes were already under 
examination \cite{Bot07,Buy13,lorente} with the statistical models. 
We plan to pursue it in the forthcoming papers because these processes are 
universal and can take place 
not only in ion reactions. Below we demonstrate in detail the results 
concerning  the hot primary hyper-residues' production, obtained within 
DCM and UrQMD+CB approaches. 

In Figs.~6, 7, and 8 we show the yields of spectator residues after the 
capture of 1, 2, and 3 $\Lambda$ hyperons. The collisions of light, 
medium, and heavy nuclei were considered in the large range of the 
projectile energies. The DCM calculations were performed with the potential 
capture criterion. The yields are presented in millibarns for the 
future convenient comparisons with experiments. Some curves are scaled 
with the factors shown in brackets on the figures. As was previously reported 
there is a trend of the yield's saturation at high energy \cite{Bot13}. 
The present calculations with large statistics demonstrate that this 
saturation trend can be still partly valid for double and even triple 
hyper-residues if the energy is well above the threshold. This confirms 
that the energies of $\sim$10 A~GeV are already sufficient for 
producing multi-strange hypernuclei. 
We have also investigated the influence of the hyperon capture potential 
on the hyper-fragment formation. Its formal decreasing from 30 to 15 MeV 
(for normal nuclear matter), which we consider as a maximum reasonable 
variation, leads to decreasing the hyper-residue yields by around 20$\%$ 
only. The reason is that the hyperon--nucleon cross-section increases 
very much at low energy, as assumed in the parametrization adopted 
in Ref.~\cite{Bot11}. 
Since the nuclear matter is moderately diluted after the cascade of first 
fast particles the low-energy interactions with remaining nucleons become 
more probable. Therefore, as a result of these secondary interactions 
the $\Lambda$ hyperon energy decreases very fastly in this energy domain 
and this looks as thermalization. 

\begin{figure}[tbh]
\includegraphics[width=0.6\textwidth]{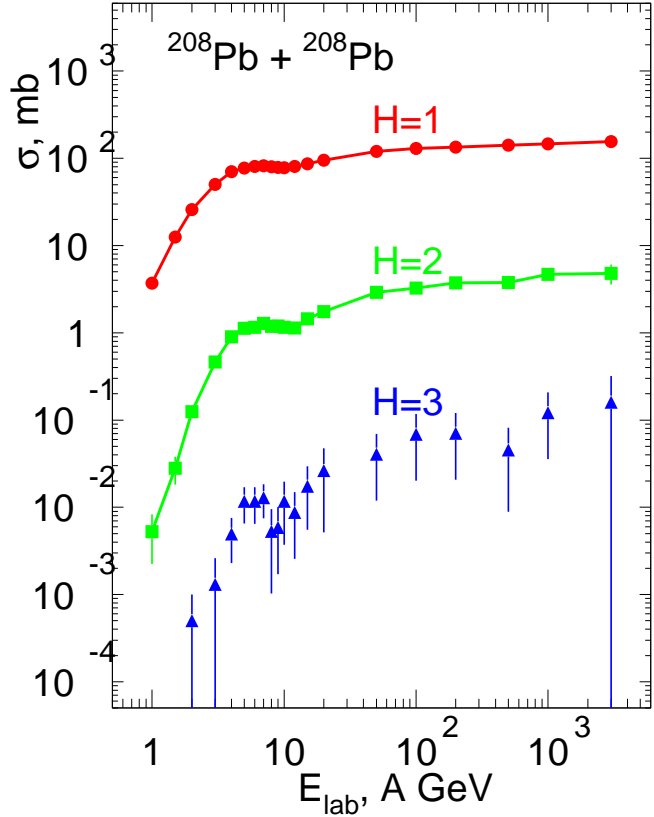}
\caption{\small{ (Color online)
Absolute yields (in mb) of the hyper-spectator residues (projectiles or 
targets) in lead on lead collisions versus the laboratory energies. 
The numbers of captured $\Lambda$ hyperons (H) are shown in the figure. 
The statistical variances ('error bars') of the performed DCM calculations 
are shown if they are larger than the size of the symbols. 
}}
\label{fig6}
\end{figure}

\begin{figure}[tbh]
\includegraphics[width=0.6\textwidth]{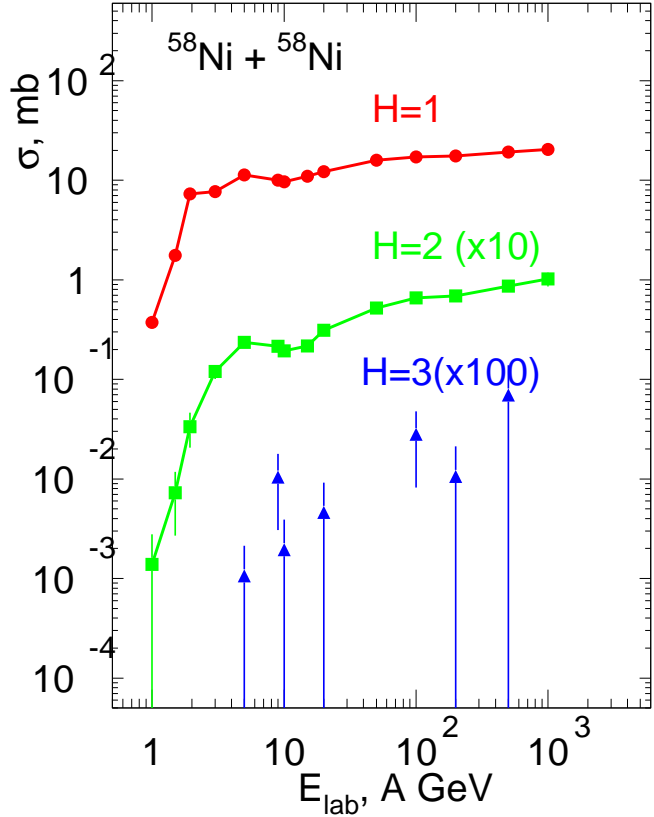}
\caption{\small{ (Color online)
Absolute yields (in mb) of the hyper-spectator residues (projectiles or 
targets) in nickel on nickel collisions. The notations are as in Fig.~6, 
besides the scaling factors given in the brackets.  
}}
\label{fig7}
\end{figure}

\begin{figure}[tbh]
\includegraphics[width=0.6\textwidth]{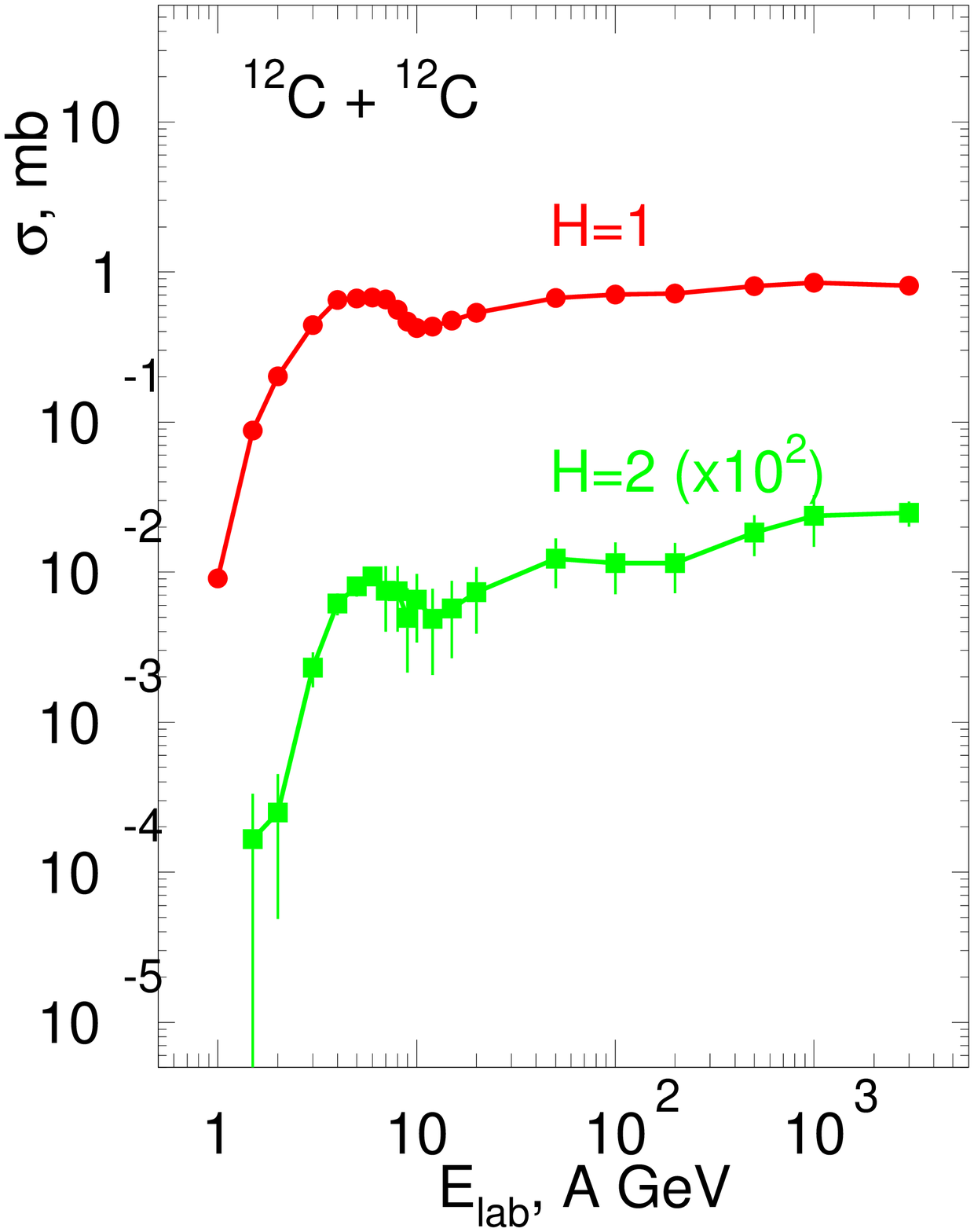}
\caption{\small{ (Color online)
Absolute yields (in mb) of the hyper-spectator residues (projectiles or 
targets) in carbon on carbon collisions. The notations are as in Figs.~6, 7. 
}}
\label{fig8}
\end{figure}

The production of hypernuclei around the threshold energy is instructive 
since it is sensitive to the properties of particles inside the nuclear 
matter of colliding nuclei, including the nucleon correlations. 
It is also practical since it can facilitate the experimental identification 
of hypernuclei, as we have mentioned in Introduction. 
To verify the predictions of the transport approaches at these energies 
we have also performed UrQMD calculations with a large statistics 
for lead on lead collisions at the energies of 1, 1.5 and 2.0 A GeV. 
With this model we can use the coalescence capture criterion and 
estimate the influence of various coalescence parameters. The procedure was 
described in detail in our previous work \cite{Bot15} and concerns both 
the relative coordinates and velocities between the coalescent nucleons. 
In particular, the 
velocity coalescence parameter $v_c \approx 0.1$ should correspond to the 
formation of lightest clusters in the ground states. While the parameter 
$v_c \approx 0.22$, encloses the nucleons with velocities close to the 
Fermi motion in nuclei. This large parameter 
is also consistent with the momenta of 
hyperons absorbed by big spectators \cite{Bot11}, therefore, it 
should be more realistic to describe the formation of larger nuclei 
which are expected for the target and projectile residues. 

As a result, the UrQMD+CB calculations for $v_c = 0.22$ predict the 
following cross-sections for producing the single hyper-residues: 
0.35 mb at 1 A GeV, 2.4 mb at 1.5 A GeV, and 9.0 mb at 2 A GeV. 
By comparing it with the DCM presented in the fig.~6, one can evaluate 
the difference between DCM and UrQMD results. 
The artificial reducing of $v_c$ to 0.1 leads to decreasing the yields to 
0.1 mb, 0.75 mb, and 2.6 mb, respectively. However, it is a clearly 
underestimated case since such small parameters are typical for lightest 
clusters (A$\loo$4), not for heavy residues. 
For the capture of two hyperons by the spectator residues in the more 
realistic $v_c = 0.22$ case we have got the following cross sections: 
0.0013 mb, 0.011 mb and 0.045 mb for 1.0, 1.5 and 2.0 A GeV, correspondently. 
One can say that various models may lead to the deviations of up to one 
order at the very low subthreshold beam energy (1 A GeV), and the difference 
becomes smaller, approximately the factor two-three, at 2 A GeV. 
At high energy, as was discussed in 
the previous works \cite{Bot11,Bot15}, the deviations in predictions of 
transport models are not more than the factor two. 

We have analyzed that 
by increasing the coalescence parameters one may try to simulate 
the effect of the spectator nucleon density fluctuations within the 
coalescence picture and increase the capture effectively. However, the main 
discrepancy between the models 
comes from the difference in the hyperon production: The yield of 
hyperons integrated over all impact parameters 
in DCM is nearly 4 times larger than in UrQMD for Pb + Pb collisions 
at 1 A GeV. This discrepancy comes from the different parametrizations 
for strangeness production and particle rescattering at low energy. 
It depends also on the effective masses and potentials of particles in 
medium. The Fermi motion of nucleons may allow for high momentum components, 
that is very important in subthreshold reactions. All these phenomena, 
which are not very crucial at very high energy, are treated in the models in 
different ways. The lack of the experimental data on low energy particles 
in subthreshold heavy-ion collisions is the main obstacle for the adequate 
adjustment of the models. 
However, we think that the presented results on the production of 
hypernuclei in the subthreshold region is a reasonable guide-line for 
their future experimental studies. Moreover, we believe that the experimental 
determination of the yields of spectator heavy hypernuclei, for example, 
by measuring remnants of the hyper-fission, may provide additional 
opportunities for the better description of the strangeness mechanisms 
inside nuclear matter at 
such low energies. This can also put the important constraint on interaction 
of hyperons in medium, since slow hyperons can be captured by 
the residues. 

\begin{figure}[tbh]
\includegraphics[width=0.6\textwidth]{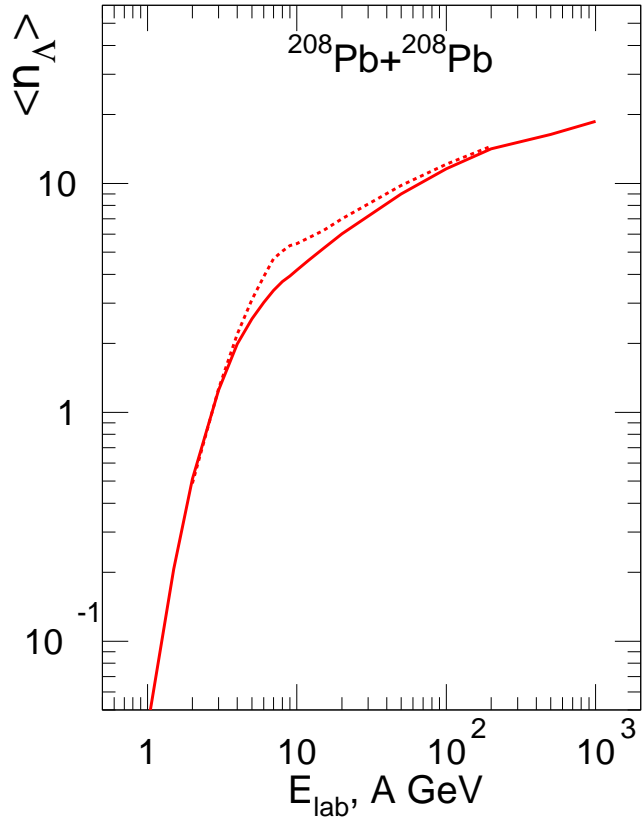}
\caption{\small{ (Color online)
The average numbers (per event) of the $\Lambda$-hyperon production in lead 
on lead collisions versus the laboratory energies. The DCM calculations are 
integrated over all impact parameters and are performed under the standard 
assumption on the smooth transition between the low-energy to high-energy 
elementary hadron interactions (solid line), and by assuming that the 
employed  
low-energy cross-section parametrizations can also be applied at the energies 
which are higher approximately by two GeV (dashed line). 
}}
\label{fig9}
\end{figure}

\begin{figure}[tbh]
\includegraphics[width=0.6\textwidth]{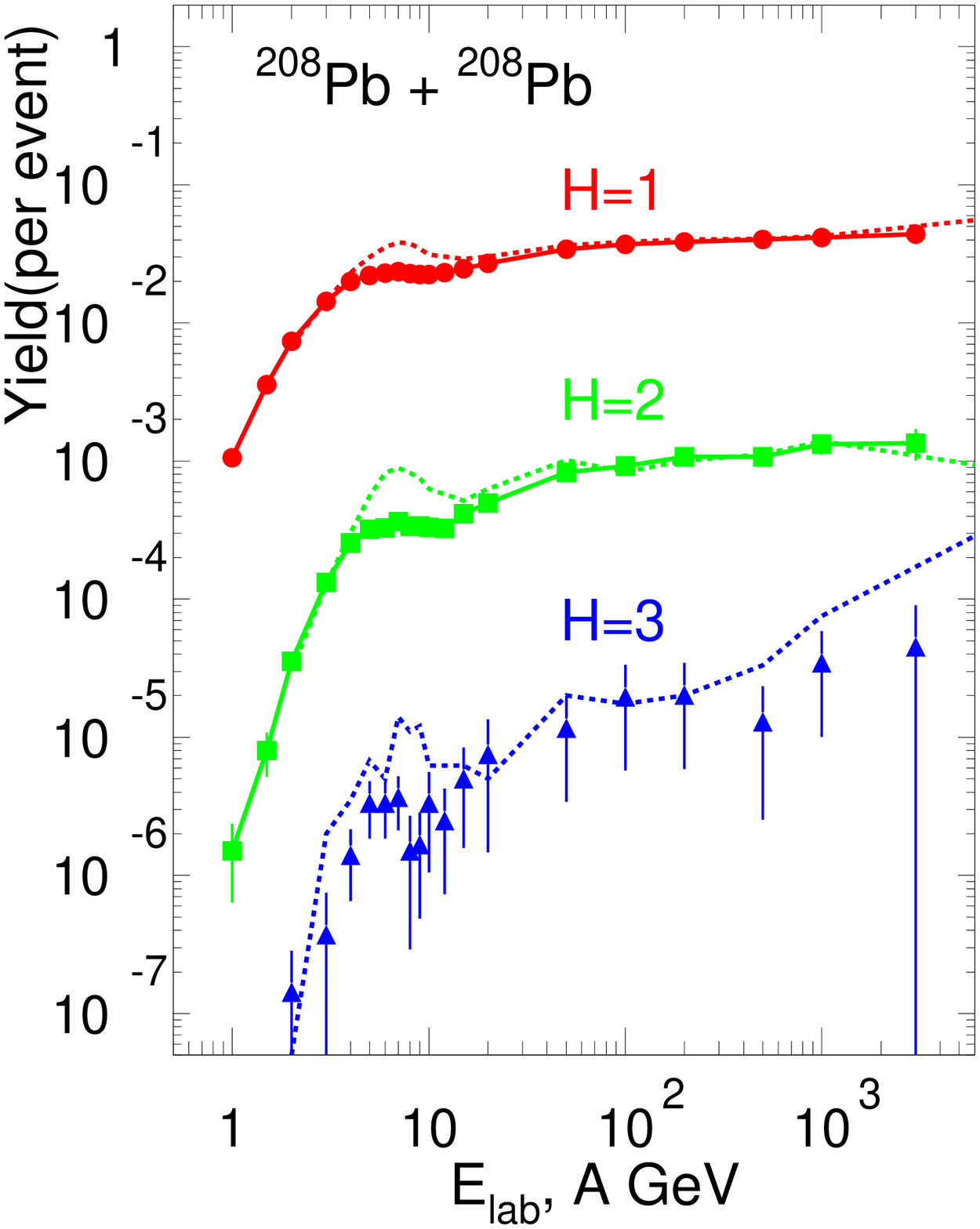}
\caption{\small{ (Color online)
Relative yields (per one inelastic event) of the hyper-spectator residues 
(projectiles or targets) in lead on lead collisions versus the laboratory 
energies. Solid lines and symbols are the DCM standard calculations. 
Dashed lines are the calculations with an alternative assumption for 
the $\Lambda$-hyperon formation shown by the dashed line in Fig.~9. 
Other notations are as in Fig.~6. 
}}
\label{fig10}
\end{figure}

It is especially instructive that the excitation functions for 
multi-strange hypernuclear residues (figs.~6--8) 
have the same saturation-like behavior. 
The probability for the formation of residues with one additional captured 
hyperon decreases by two order of magnitude in the collisions of heavy 
nuclei. This difference increases up to four order of magnitude for the 
very light nuclei. Actually, our predicted yields of the hyper-residues 
may be parameterized in the wide mass range (from carbon to lead) and used 
for preparing the corresponding measurements. 
The reason of the decreasing of the hyper-residues yields for smaller 
colliding nuclei is just that less hyperons are 
produced in the collision events. Still these cross sections are sufficient 
for the systematic investigation of hypernuclei. Moreover, in some cases 
the light colliding nuclei have advantages: The background conditions are 
better for experimental identification of hypernuclei, and their mesonic 
decay channels gives a chance to use the invariant mass methods well 
established in hypernuclei studies \cite{saito-new}. 
In relativistic ion reactions these correlations were investigated 
theoretically too (Ref.~\cite{Bot13}). 
Therefore, the first experiments may take place on light nuclei 
\cite{Rap13}. One can see that in the case of carbon collisions (fig.~8) 
at the beam energy around 2 A GeV even double light hyper-residues can 
be produced with the cross-section 
approximately $\sim$1 nanonbarn. Taking into account the high intensities 
of future accelerators (e.g., the planned rate is nearly 10$^{12}$ per 
second for 
the FAIR beam \cite{frs}) it is sufficient for starting a rich hypernuclear 
program. 

It is also instructive to understand within the transport models how the 
variation of the hyperon production parameters can be seen in the production 
of hyper-residues. In fig.~9 we show the prediction of the $\Lambda$ 
hyperon yields by using different parametrizations inside DCM: 
The solid line is the standard assumption on the transition from the 
well-known low-energy regime to the high energy elementary interactions 
described within the QGSM \cite{Ton90}. The dashed line presents another 
procedure for fitting these two limits. We note that up to now there are no 
sufficient experimental data available for comparison to make unambiguous 
conclusion about the correct excitation function of the hyperons. A small 
variation of the $\Lambda$ yields at the energies slightly below 10 A GeV 
may result in a specific feature of the hyper-residue yields: We see from 
Fig.~10 that if we involve an alternative DCM parametrization, leading to 
the dashed line in Fig.~9, one can get 
even local maxima of these yields at the corresponding energies. This is a 
consequence of that the secondary interactions, as well as the hyperon 
capture in the potential 
well, are very sensitive to the details of the hyperon origin and energy. 
Such kind of behaviour of the excitation functions could be a very 
instructive experimental signature complementary to measuring high-energy 
spectra of strange particles. It may compensate partly the lack of the 
low-energy kaon data, since predominantly low-energy hyperons can be 
absorbed inside nuclei. 

\begin{figure}[tbh]
\includegraphics[width=0.6\textwidth]{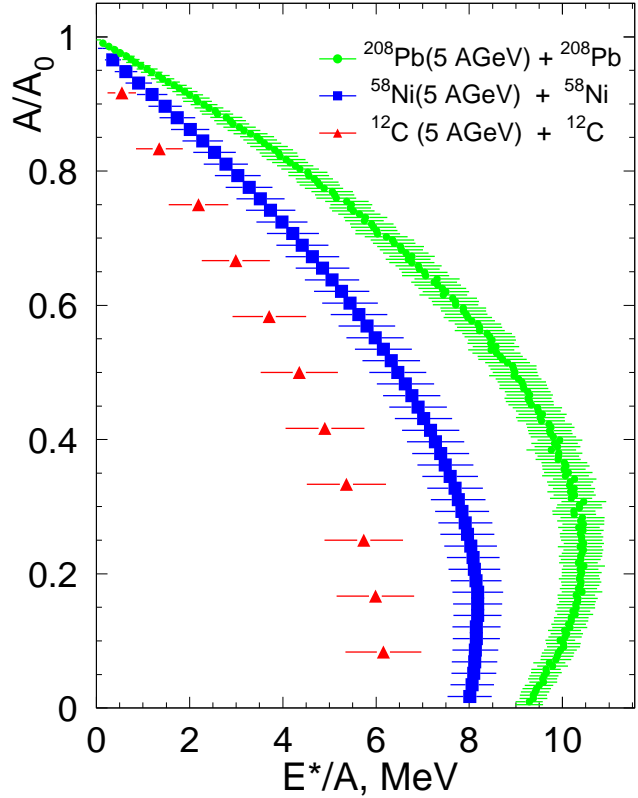}
\caption{\small{ (Color online)
The average mass numbers of the produced residues, divided by mass number 
of the corresponding target and projectile, versus their excitation energy 
per nucleon. The 'error bars' by the symbols give the standard deviations 
of the DCM calculations. Reactions and laboratory energy of collision 
nuclei are indicated in the figure by symbols. 
}}
\label{fig11}
\end{figure}

For the following description of the reaction processes it is important 
to have information about the properties of the spectator hyper-residues. 
Since the masses and excitation energies of nuclei in such intermediate 
states serve as input for the statistical de-excitation models. This problem 
was under intensive studies in relativistic heavy-ion collisions leading 
to the normal nuclear fragmentation. For example, as was experimentally 
established, there is a special correlation between a residue masses and 
the excitations \cite{Bon95,Xi97,Ogu11,Ima15} 
which results in an universal fragmentation picture. 
The hyper-residue masses have been demonstrated in our previous works, 
see, e.g., fig.~7 in Ref.~\cite{Bot11} and fig.~2 in Ref.~\cite{Bot15}: 
They range from the small mass numbers to the ones close to the target and 
projectile. This allows for investigating a very broad distribution of 
hypernuclei (in mass and isospin) in the same collisions. The 
connection between masses and excitations obtained in the DCM calculations, 
for the first time, 
is shown in Fig.~11. One can see that by increasing the collision violence 
the more nucleons are lost and the more 
excitation energy is deposited per nucleons. However, there is a saturation 
trend for the excitation, so the excitation energies which exceed essentially 
the nuclear binding energies are not realized in thermalized residues. 
We have checked that in our calculations this trend is fully 
consistent with the previous analysis of experimental data and it remains 
valid for collisions in the wide range of relativistic energies available 
at GSI/FAIR and other accelerators. 
Actually, such excited residual nuclei will decay in the fast 
multifragmentation/break-up and/or sequential evaporation/fission processes 
leading to cold hyper- and normal nuclei \cite{Bon95,Bot07,Bot13}. 
In the case of such de-excitation the captured hyperons 
will be predominantly concentrated in the biggest final fragments 
because of the considerable hyperon binding energy, see Ref.~ \cite{Buy13}. 

As the last step, for identification of hypernuclei, 
the correlation measurements (of pions, baryons and fragments) are the 
most promising tool for future research in this field 
\cite{star,alice,saito-new,Bot13,Arm93,Ohm97}. 
Besides identifying hypermatter, 
the correlations can reveal the hypernuclei properties. 
For example, by detecting the momenta of the decay products one can find 
the life-time of the hypernuclei and their binding energies. By analyzing 
the decay of free $\Lambda$ hyperons and hypernuclei in the same events one 
can investigate the unbound hyperon states in double hypernuclei. It is 
crucial for 
constraining the hyperon interaction in matter and determining the 
properties of hypermatter at low temperatures. 

\section{Conclusion}

We conclude that the spectator region in relativistic collisions of hadron 
and ion with ions can be a very promising source of hyper-fragments. 
We point that the general mechanism of such reactions leading to 
fragmentation and multifragmentation is well established for normal nuclear 
processes. Hyperons are also participating in such a process 
because the hyperon-nucleon interaction is of the same order as 
the nucleon-nucleon one. The 
primary produced hypermatter is relatively cold (the expected temperature 
of the spectator residues is not higher than T$\sim$5--7 MeV), therefore, 
large hypernuclei can be 
produced in comparison with the central collisions. A great variety 
of hypernuclei of all masses and in a wide range of isospin can be formed, 
that is similar to the phenomena existing in normal nuclei. 
Systematic investigations of strange and, especially, 
multi-strange hypernuclei can be 
naturally performed in these reactions. We have demonstrated the quantitative 
estimates for the yields of such primary hyperfragments. 
The calculations are partly confirmed by a rather good description of 
experimental data available on the strangeness production. 
We have also investigated the 
theoretical uncertainties of the predictions by considering the variations of 
the model parameters, and different transport models. 
As we have found these 
uncertainties can be related to the treatment of the strange 
particle interaction in medium at low energy, and this opens an complementary 
way for experimental investigation of such processes. 
From the current experiments we know that the values of the hypernuclei 
yields obtained within our approach are sufficient for the systematic 
experimental measurements. 
Moreover, our predictions of the yields can be naturally extended 
for the whole mass and energy range available for targets/projectiles in 
future experiments. 
The saturation of the hypernuclei production  
at high laboratory energies indicate that high intensities of the accelerators 
and a more sophisticated detection technique are more important for this 
purpose than the ultra-high colliding energies. 

In this respect it is encouraging that the residues of ions and their decay 
products with energies from 1--2 A GeV (i.e., around the hypernuclear 
threshold) and up to 10--15 A GeV can be effectively studied with the 
modern experimental installations, like FRS/Super-FRS and CBM at GSI and FAIR. 
These experiments are in preparation. New exotic hypernuclei 
can be investigated in such reactions, and new methods of their 
determination (e.g., by using many-particle correlations) 
can be applied, which may give advantages over the traditional hypernuclear 
studies.

\begin{acknowledgments}
A.S.B. acknowledges the support of BMBF. Also we acknowledge a partial 
support of GSI and the Research Infrastructure 
Integrating Activity Study of Strongly Interacting Matter HadronPhysics3 
under the 7th Framework Programme of EU (SPHERE network). A.S.B. and K.K.G. 
thank the Frankfurt Institute for Advanced Studies J.W. Goethe University 
for hospitality. 
\end{acknowledgments}


\begin{thebibliography}{99}

\bibitem{Ban90} H. Bando, T. Mottle, and J. Zofka, Int. J. Mod. Phys. 
{\bf A5}, 4021 (1990).

\bibitem{Sch93} J. Schaffner, C.B. Dover, A. Gal, C. Greiner, and 
H. Stoecker, Phys. Rev. Lett. {\bf 71}, 1328 (1993).

\bibitem{Gre96} W. Greiner, J. Mod. Phys. {\bf E5}, 1 (1996).

\bibitem{Has06} O. Hashimoto, H. Tamura, Prog. Part. Nucl. Phys. {\bf 57}, 
564 (2006).

\bibitem{Sch08} J. Schaffner-Bielich, Nucl. Phys. A {\bf 804}, 309 (2008).

\bibitem{Gal12} Special issue on {\em Progress in Strangeness Nuclear Physics},
Edt. A. Gal, O Hashimoto and J. Pochodzalla, 
Nucl. Phys. A {\bf 881}, 1-338 (2012).

\bibitem{Buy13} N. Buyukcizmeci, A.S. Botvina, J. Pochodzalla, and 
M. Bleicher, Phys. Rev. C {\bf 88}, 014611 (2013).

\bibitem{Hel14} T. Hell and W. Weise, Phys. Rev. C {\bf 90}, 045801 (2014).



\bibitem{star} The STAR collaboration, Science {\bf 328}, 58 (2010).

\bibitem{alice} B. D\"onigus {\em et al.} (ALICE collaboration), 
Nucl. Phys. A{\bf 904-905}, 547c (2013).

\bibitem{panda} The PANDA collaboration, http://www-panda.gsi.de ;
and arXiv:physics/0701090.

\bibitem{Hades} G. Agakishiev {\em et al.}, arXiv:1310.6198 (2013).

\bibitem{Vas15} I. Vassiliev for CBM collaboration. 
Hypernuclei program at the CBM experiment. HYP2015, 
http://indico2.riken.jp/indico/contributionListDisplay.py?confId=2002

\bibitem{saito-new} T.R. Saito {\em et al.} (HypHI collaboration),
Nucl. Phys. A {\bf 881}, 218 (2012); 
C.~Rappold {\em et al.}, Phys. Rev. C {\bf 88}, 041001 (R) (2013). 

\bibitem{super-frs}  https://indico.gsi.de/event/superfrs3  (access to pdf
files via timetable and key 'walldorf').

\bibitem{Rap13} C.Rappold, T.R. Saito, and C. Scheidenberger.
Simulation Study of the Production of Exotic Hypernuclei at the Super-FRS
(at GSI Scientific report 2012), GSI Report 2013-1, 176 p. (2013).
http://repository.gsi.de/record/52079 .

\bibitem{nica} NICA White Paper, 
http://theor.jinr.ru/twiki-cgi/view/NICA/WebHome ;
http://nica.jinr.ru/files/BM@N .

\bibitem{Dan53} M. Danysz and J. Pniewski, Philos. Mag. {\bf 44}, 348 (1953).

\bibitem{Bon95} J.P. Bondorf, A.S. Botvina, A.S. Iljinov, I.N. Mishustin, 
and K. Sneppen,  Phys. Rep. {\bf 257}, 133 (1995).

\bibitem{Xi97} H. Xi  {\em et al.}, Z. Phys. A {\bf 359}, 397 (1997).

\bibitem{Sch01} R.P. Scharenberg  {\em et al.},  Phys. Rev. C {\bf 64}, 
 054602 (2001).

\bibitem{Ogu11} R. Ogul {\em et al.}, Phys. Rev. C {\bf 83}, 024608 (2011).

\bibitem{Bot07} A.S. Botvina and J. Pochodzalla, Phys. Rev. C {\bf 76}, 
 024909 (2007).

\bibitem{Das09} S.Das Gupta, Nucl. Phys. A {\bf 822}, 41 (2009).

\bibitem{Bot11}
A.S.~Botvina, K.K.~Gudima, J.~Steinheimer, M.~Bleicher, I.N.~Mishustin,
Phys. Rev. C {\bf 84}, 064904 (2011).

\bibitem{Bot13} A.S.~Botvina, K.K.~Gudima, J.~Pochodzalla,
Phys. Rev. C {\bf 88}, 054605 (2013).

\bibitem{Bot15} A.S. Botvina {\em et al.}, Phys. Lett. B {\bf 742}, 7 (2015). 

\bibitem{aumann} Th. Aumann, Progr. Part. Nucl. Phys. {\bf 59}, 3 (2007). 

\bibitem{frs} H. Geissel {\em et al.}, 
Nucl. Inst. Meth. Phys. Res. B {\bf 204}, 71 (2003). 

\bibitem{ygma-nufra} Y.-G. Ma (for STAR/RHIC collaboration), talk at
NUFRA2013 conference, Kemer, Turkey, 2013,
http://fias.uni-frankfurt.de/historical/nufra2013/.

\bibitem{camerini-nufra} P. Camerini (for ALICE/LHC collaboration), talk at
NUFRA2013 conference, Kemer, Turkey, 2013, 
http://fias.uni-frankfurt.de/historical/nufra2013/.

\bibitem{And11} A. Andronic, P. Braun-Munzinger, J. Stachel, H. St\"ocker,
 Phys.Lett. B {\bf 697}, 203 (2011).

\bibitem{Ste12} J.~Steinheimer, K.~Gudima, A.~Botvina, I.~Mishustin, 
M.~Bleicher,
H.~St\"ocker, Phys. Lett. B {\bf 714}, 85 (2012).

\bibitem{Wak88} M. Wakai, H. Bando and M. Sano, Phys. Rev. C {\bf 38},
 748 (1988).

\bibitem{Cas95} Z.Rudy, W.Cassing {\em et al.}, Z. Phys. A {\bf 351},
 217 (1995).

\bibitem{giessen} Th. Gaitanos, H. Lenske, and U. Mosel,
Phys. Lett. B {\bf 675}, 297 (2009).

\bibitem{Arm93} T.A. Armstrong {\em et al.},
Phys. Rev. C {\bf 47}, 1957 (1993).

\bibitem{Ohm97} H. Ohm {\em et al.},
Phys. Rev. C {\bf 55}, 3062 (1997).

\bibitem{Poc97} J. Pochodzalla, Prog. Part. Nucl. Phys. {\bf 39}, 443 (1997).

\bibitem{Ton90} V.D.~Toneev, N.S.~Amelin, K.K.~Gudima, S.Yu.~Sivoklokov, 
Nucl. Phys. A {\bf 519}, 463c (1990). 

\bibitem{Bas98} S.A.~Bass {\em et al.}, Prog. Part. Nucl. Phys. {\bf 41}, 
 225 (1998).

\bibitem{Ble99} M.~Bleicher {\it et al.}, J. Phys. G {\bf 25}, 1859 (1999). 

\bibitem{Cas08} W. Cassing and E.L. Bratkovskaya,
Phys. Rev. C {\bf 78}, 034919 (2008).

\bibitem{lorente} A.S. Lorente, A.S. Botvina, and J. Pochodzalla, 
Phys. Lett. B {\bf 697}, 222 (2011). 

\bibitem{Bra04} E.L. Bratkovskaya {\em et al.}, 
Phys. Rev. C {\bf 69}, 054907 (2004). 

\bibitem{Har12} C. Hartnack {\em et al.}, 
Phys. Rep. {\bf 510}, 119 (2012). 

\bibitem{Fuc01} C. Fuchs {\em et al.}, Phys. Rev. Lett. {\bf 86}, 1974 (2001).


\bibitem{FOPI04} X. Lopez (for FOPI collaboration), Prog. Part. Nucl. Phys. 
{\bf 53}, 149 (2004). 

\bibitem{PRL01} C. Sturm {\em et al.}, Phys. Rev. Lett. {\bf 86}, 39 (2001). 

\bibitem{PRC07} A.~F\"orster {\em et al.}, Phys. Rev. C {\bf 75}, 024906 
(2007). 

\bibitem{PRL06} W. Scheinast {\em et al.}, Phys. Rev. Lett. {\bf 96}, 072301 
(2006). 

\bibitem{Gai14} T. Gaitanos and H. Lenske, 
Phys. Lett. B {\bf 737}, 256 (2014). 

\bibitem{LeFev16} A. Le Fevre {\em et al.}, J. Phys. Conf. Ser. {\bf 668}, 
012021 (2016).

\bibitem{Ima15} H. Imal {\em et al.}, Phys. Rev. C {\bf 91}, 034605 (2015). 

\end{thebibliography}
\end{document}